\documentclass[a4paper,11pt]{article}
\usepackage{pos}

\usepackage{caption}
\usepackage{subcaption}

\title{DUNE: science and status}
\author*{Francisco Mart\'{i}nez L\'{o}pez}
\onbehalf{for the DUNE collaboration}

\affiliation[]{School of Physical and Chemical Sciences, Queen Mary University of London,\\
London E1 4NS, United Kingdom}

\emailAdd{f.martinezlopez@qmul.ac.uk}

\abstract{The Deep Underground Neutrino Experiment (DUNE) is a next-generation long-baseline neutrino oscillation experiment. Its primary goal is the determination of the neutrino mass hierarchy and the CP-violating phase. The DUNE physics program also includes the detection of astrophysical neutrinos and the search for beyond the Standard Model phenomena, such as nucleon decays. DUNE will consist of a near detector complex placed at Fermilab, several hundred meters downstream of the neutrino production point, and 17-kton Liquid Argon Time Projection Chamber (LArTPC) far detector modules to be built in the Sanford Underground Research Facility (SURF), approximately 1.5 km underground and 1300 km away. The detectors will be exposed to a wide-band neutrino beam generated by a 1.2 MW proton beam, with a planned upgrade to 2.4 MW. Two prototypes of the FD technology, the ProtoDUNE 700 ton LArTPCs, have been operated at CERN for over 2 years, and have been recently optimized to take new data in 2024-2025. Additionally, the 2x2 Demonstrator, a prototype of the LAr component of the near detector, has recently started operations in the NuMI beam at Fermilab. This talk will present the science programme, as well as recent progress, of DUNE and its different prototyping efforts.}

\FullConference{The 43rd International Symposium on Physics in Collision (PIC2024)\\
 22-25 October 2024\\
Athens, Greece\\}

\begin{document}
\maketitle

\begin{figure}[t]
	\centering
	\includegraphics[width=0.85\linewidth]{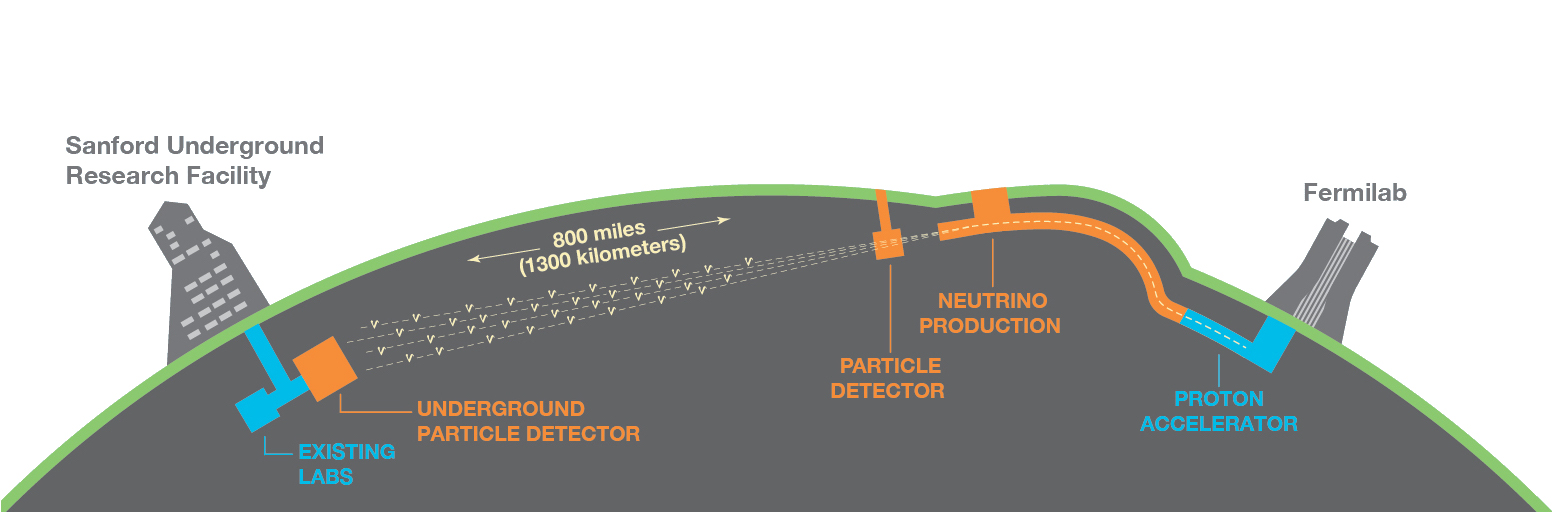}
	\caption{Schematic diagram of the DUNE experiment and the LBNF beamline \cite{DUNE2020TDR1}.}
	\label{fig:dune}
\end{figure}

The Deep Underground Neutrino Experiment (DUNE) is a next generation long-baseline neutrino oscillation experiment. The main physics goals of DUNE are \cite{DUNE2020TDR1}:
\begin{itemize}
	\item measure the neutrino mass hierarchy, the amount of CP violation in the leptonic sector and the $\theta_{23}$ octant,
	\item detect low energy neutrino events, like neutrinos from supernova bursts or the Sun, and
	\item search for proton decay and other beyond the standard model phenomena.
\end{itemize}

The design of DUNE has been tailored with these goals in mind. It will consist of two neutrino detectors. A near detector (ND) complex will be placed at Fermilab, $574~\mathrm{m}$ downstream of the neutrino production point, whereas a larger far detector (FD) will be built in the Sandford Underground Research Facility (SURF), South Dakota, approximately $1300~\mathrm{km}$ away. Figure \ref{fig:dune} shows a simplified diagram with the various components of DUNE (not to scale).

\section{Physics programme}

As noted in the literature (see for instance Ref. \cite{deSalas2020} for a review), the parameter space of the neutrino oscillation phenomena within the three-flavour picture is quite constrained by current experimental data. However, there are still crucial open questions, like the mass ordering, the value of $\delta_{CP}$ or the $\theta_{23}$ octant. One of the main goals of DUNE is to determine precisely the values of these parameters \cite{DUNE2020TDR2}.

\begin{figure}[t]
	\begin{subfigure}{0.49\textwidth}
		\centering
		\includegraphics[width=.99\linewidth]{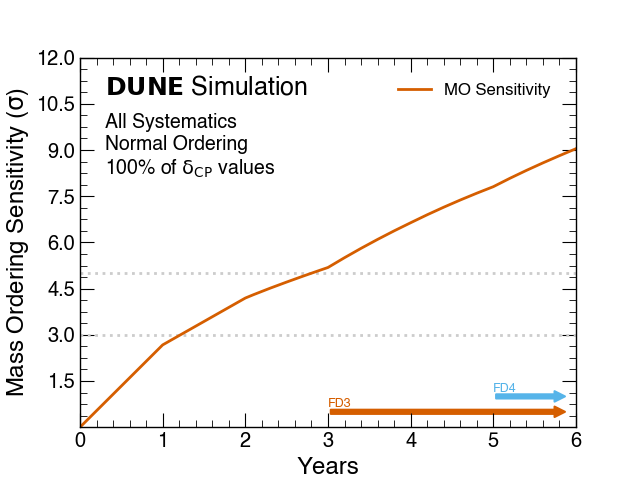}
	\end{subfigure}
	\begin{subfigure}{0.49\textwidth}
		\centering
		\includegraphics[width=.99\linewidth]{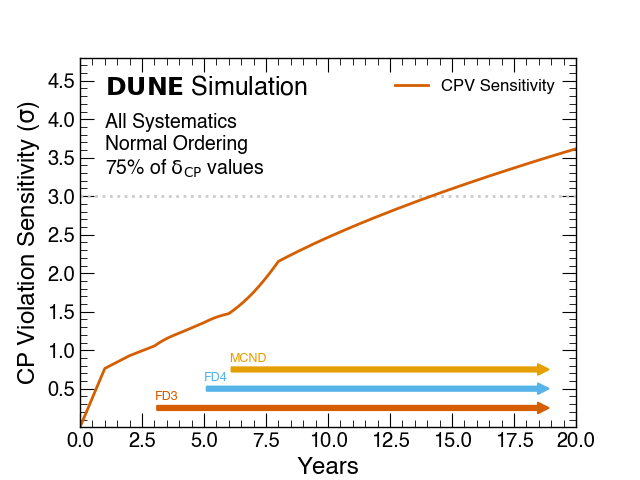}
	\end{subfigure}
	\caption{Significance of the DUNE determination of the neutrino mass ordering (left panel) and CP-violation (right panel) for $100\%$ and $75\%$ of possible true $\delta_{\mathrm{CP}}$ values, respectively, as a function of the exposure in years \cite{DUNE2020TDR2}.}
	\label{fig:oscillation_sensitivity}
\end{figure}

To address these questions DUNE can look to the subdominant oscillation channel $\nu_{\mu} \rightarrow \nu_{e}$ ($\bar{\nu}_{\mu} \rightarrow \bar{\nu}_{e}$) and study the energy dependence of the $\nu_{e}$ ($\bar{\nu}_{e}$) appearance probability. When we focus on the antineutrino channel $\bar{\nu}_{\mu} \rightarrow \bar{\nu}_{e}$ there is a change in the sign of $\delta_{CP}$, thus introducing CP-violation. Moreover, due to the fact that there are no positrons in the composition of the Earth, there is a sign difference for the matter effect contribution when looking to the antineutrino channel. This asymmetry is proportional to the baseline length $L$ and is sensitive to the sign of $\Delta m^{2}_{31}$, and thus to the neutrino mass ordering. Figure \ref{fig:oscillation_sensitivity} shows the sensitivity of DUNE to the mass ordering and CP-violation as a function of time, assuming the current staging scenario.

DUNE also aims to detect neutrinos originated in supernovae explosions, in what is called a supernova neutrino burst (SNB). These neutrinos carry with them information about the core-collapse process, from the progenitor to the explosion and the remnant; but also may have information about new exotic physics. Although these are quite rare, as the expected supernovae explosion events are about one every few decades for our galaxy and Andromeda, the long lifetime of the experiment makes it reasonable to expect some.

Nowadays the main sensitivity to SNB of most experiments is to the $\bar{\nu}_{e}$ flux through inverse beta decay. One of the advantages of DUNE is its expected sensitivity to MeV-scale $\nu_{e}$ events, since the dominant channel will be $\nu_{e}$ CC scattering. Additionally, DUNE is expected to have excellent sensitivity to $^{8}\mathrm{B}$ solar neutrinos and discovery potential for the hep solar flux. Studies are under way to understand the impact of DUNE on the solar oscillation measurements.

Moreover, due to the stringent requirements that the main physics goals set for DUNE, it will allow to perform searches for all kind of BSM physics. Among others, DUNE will be able to look for: active-sterile neutrino mixing, non-unitarity of the PMNS matrix, non-standard interactions, Lorentz and CPT violations, neutrino trident production, light-mass DM, boosted DM, and heavy neutral leptons. The reader is referred to the DUNE FD TDR Volume II \cite{DUNE2020TDR2} for a full discussion of the physics scope of DUNE.

\section{DUNE detectors}

\begin{figure}[t]
	\begin{subfigure}{0.49\textwidth}
		\centering
		\includegraphics[width=.85\linewidth]{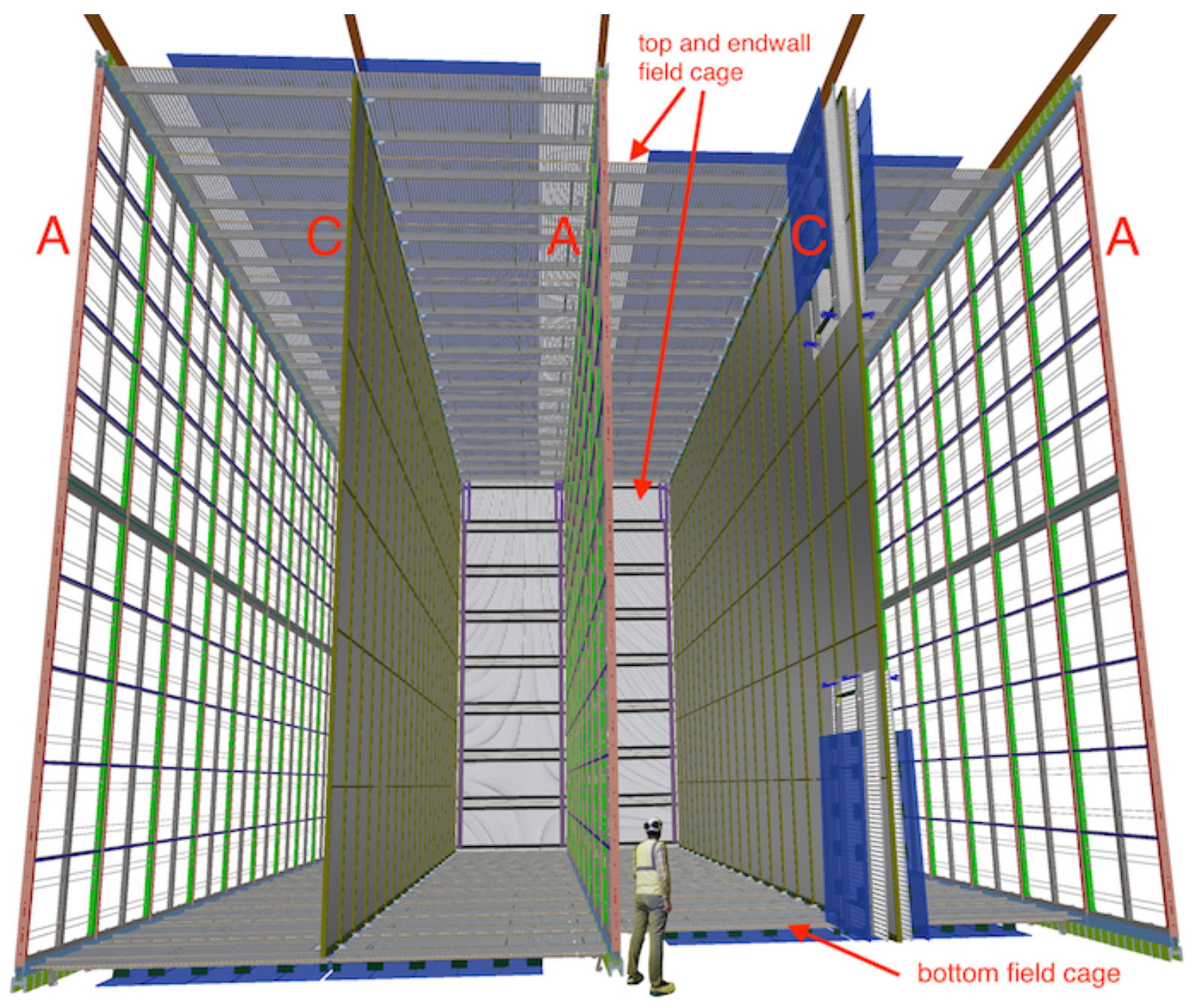}
	\end{subfigure}
	\begin{subfigure}{0.49\textwidth}
		\centering
		\includegraphics[width=.95\linewidth]{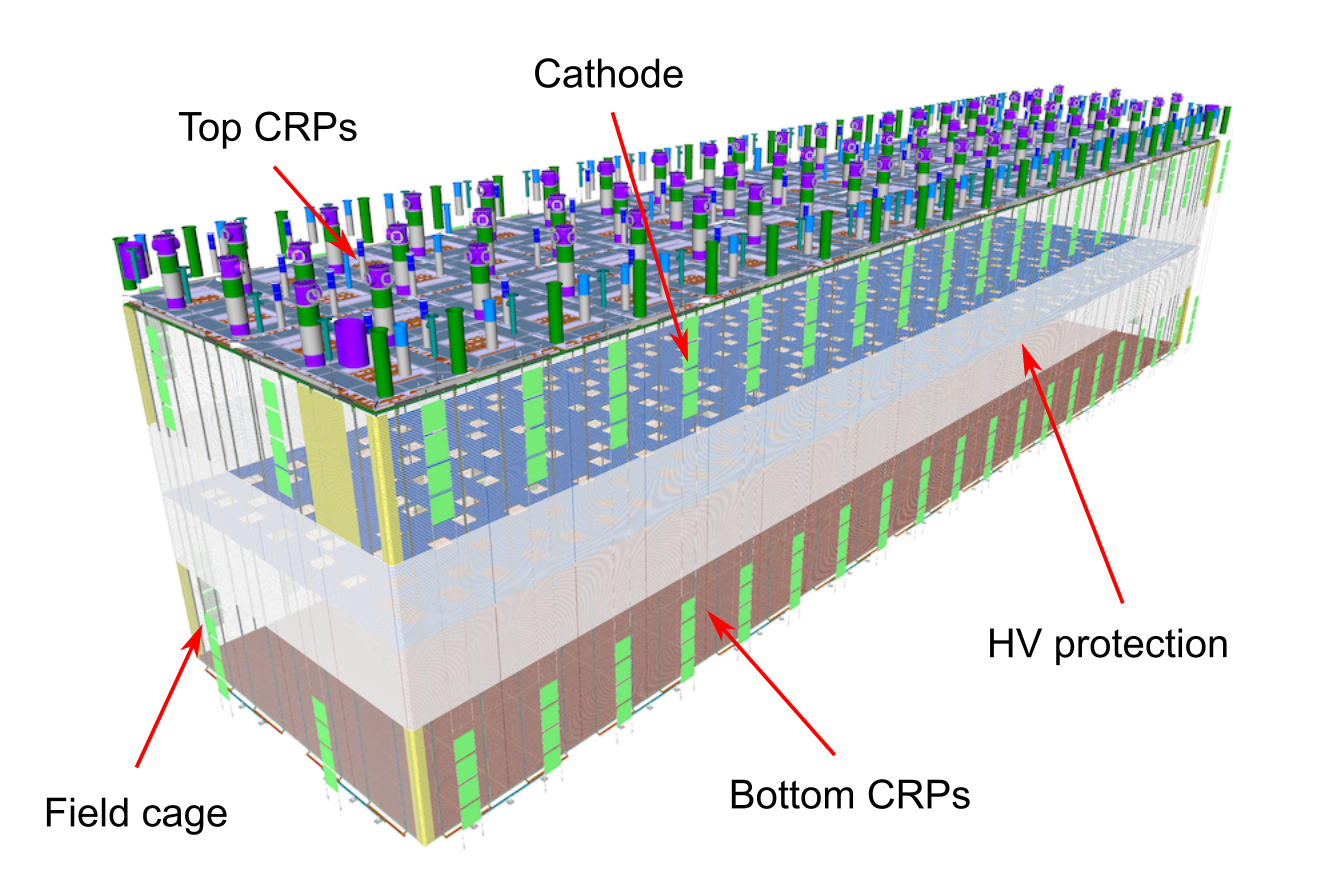}
	\end{subfigure}
	\caption{Proposed designs for the DUNE FD modules following the HD (left) and VD (right) principles. Figure taken from Refs. \cite{DUNE2020TDR1} and \cite{DUNE2024VDTDR}.}
	\label{fig:dune_far_detectors}
\end{figure}

The beam neutrinos will be provided by the Long-Baseline Neutrino Facility (LBNF) beamline, the multi-megawatt wide-band neutrino beam planned for Fermilab. It will produce neutrinos travelling in the direction of SURF, with the capability to switch between neutrino and antineutrino mode. A detailed discussion of the LBNF programme can be found in the DUNE/LBNF CDR Volume III \cite{DUNE2016CDR3}.

The technology chosen for DUNE FD is the liquid Argon time projection chamber (LArTPC). Its four modules will record neutrino interactions from the accelerator-produced beam arriving at predictable times. As it also aims at recording rare and low energy events the FD requires trigger schemes which can deal with both kinds of physics, and also maximum uptime. 

The first and third FD modules, FD-1 and FD-3, will use a Vertical Drift (VD) technology, whereas the second module, FD-2, will have a Horizontal Drift (HD) direction. The technology for the fourth module is still to be decided. The design of these modules is shown in Fig. \ref{fig:dune_far_detectors}. Detailed descriptions of the HD and VD designs can be found in the DUNE FD TDR Volume IV \cite{DUNE2020TDR4} and the DUNE FD VD TDR \cite{DUNE2024VDTDR}, respectively.

Before arriving to the FD, the neutrino beam meets the ND complex, which serves as the experiment's control. The design of the DUNE ND is mainly driven by the needs of the oscillation physics programme, as its main role is to measure the unoscillated neutrino energy spectra. From these we can predict the unoscillated spectra at the FD, which can be compared to the spectra measured at the FD to extract the oscillation parameters. Additionally, the ND has a physics programme of its own, including cross section measurements and BSM physics searches.

The DUNE ND can be divided in three main components: a LArTPC known as ND-LAr, built by optically isolated modules with a 3D pixel-based charge readout, a magnetised muon spectrometer, which in day one will be The Muon Spectrometer (TMS), and the System for on-Axis Neutrino Detection (SAND). The layout of the DUNE ND can be seen in Fig. \ref{fig:dune_nd} (left panel). The first two components of the ND will be able to move off-axis, in what is called the Precision Reaction-Independent Spectrum Measurement (PRISM) concept. More details on the purpose and design of the ND can be found in the DUNE ND CDR \cite{DUNE2021NDCDR}.

\section{DUNE Phase II}

\begin{figure}[t]
	\begin{subfigure}{0.49\textwidth}
		\centering
		\includegraphics[width=.90\linewidth]{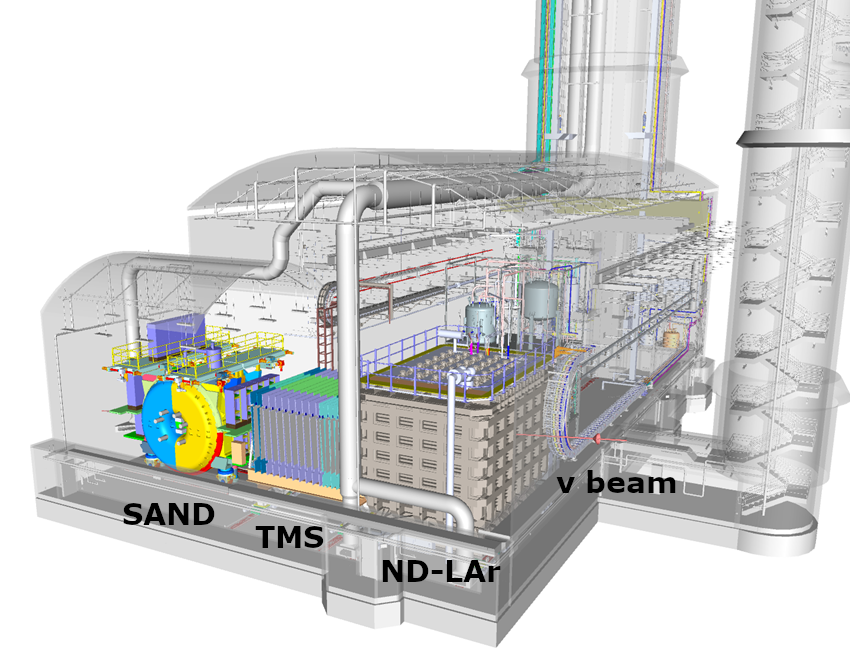}
	\end{subfigure}
	\begin{subfigure}{0.49\textwidth}
		\centering
		\includegraphics[width=.70\linewidth]{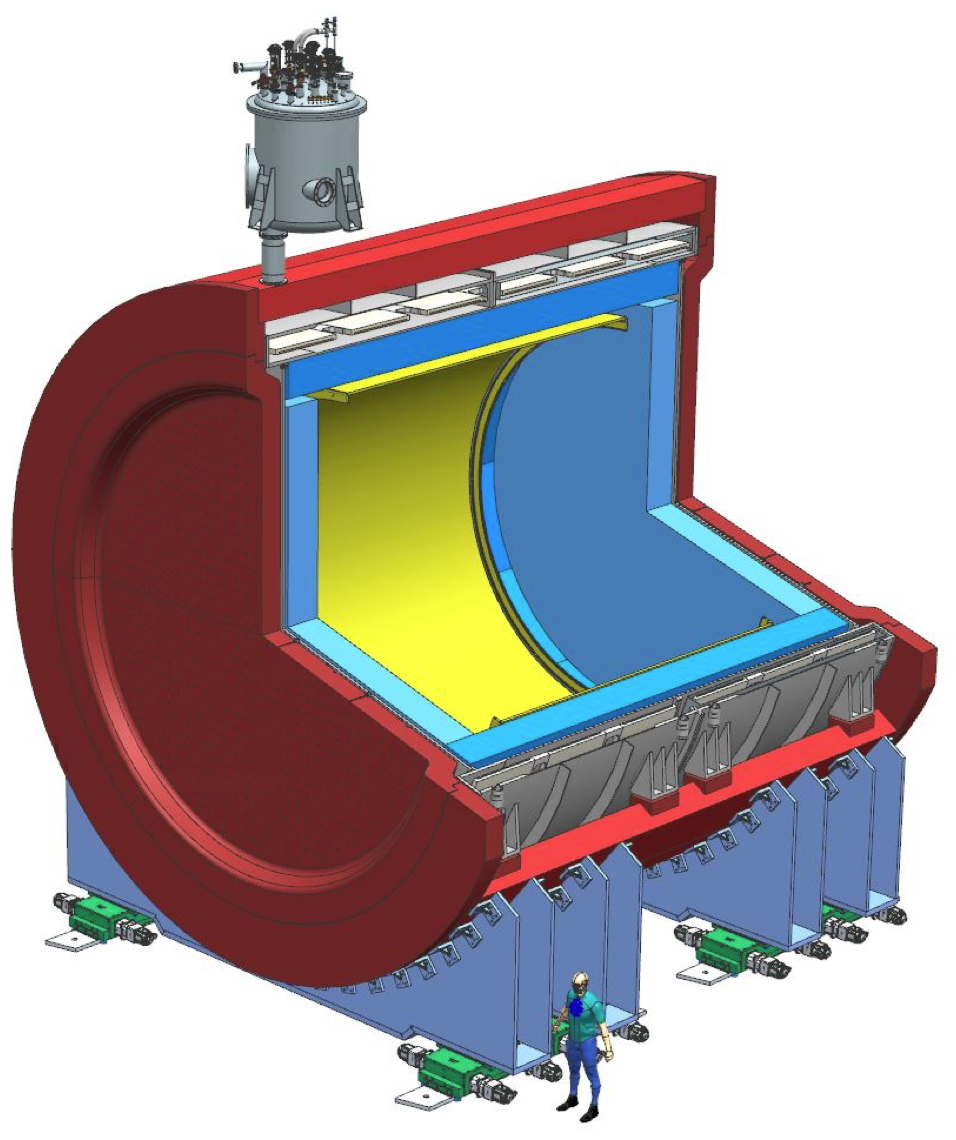}
	\end{subfigure}
	\caption{Left panel: representation of the DUNE ND hall in Phase I, showing the different subcomponents. Right panel: cross section of the ND-GAr geometry, showing the HPgTPC, ECal, and magnet. Figures adapted from Ref. \cite{DUNE2021NDCDR}.}
	\label{fig:dune_nd}
\end{figure}

DUNE is planned to be built using a staged approach consisting on two phases. Phase I consists of a FD with $50\%$ of the total fiducial mass, a reduced version of the ND complex and a $1.2~\mathrm{MW}$ proton beam. It will be sufficient to achieve some early physics goals, like the determination of the neutrino mass ordering. For its Phase II, DUNE will feature the full four FD modules, a more capable ND and a $>2~\mathrm{MW}$ proton beam. The current staging scenario assumes that Phase II is completed after 6 years of operation. For a detailed discussion on the two-phased approach the reader is referred to the DUNE Snowmass 2021 report \cite{DUNE2022Snowmass}.

VD is the baseline design for the Phase II FD modules. However, the phased construction allows the technological developments to expand the physics of DUNE (solar and supernova neutrinos, $0\nu\beta\beta$, DM, …). We are currently pursuing improvements to the light collection system for FD-3, including Aluminium Profiles with Embedded X-ARAPUCA (APEX). For FD-4 (the so-called ``Module of Opportunity'') more ambitious designs are being considered, in particular: pixelated readout, integrated charge-light readout, low background modules, Xe doping, non-LAr options, etc. A summary of the different options considered can be found in Ref. \cite{DUNE2024PhaseII}.

The current leading proposal for the Phase II ND is ND-GAr. This detector is a magnetised, high-pressure gaseous argon TPC (HPgTPC) surrounded by an electromagnetic calorimeter and a muon tagger. A cross section of its geometry can be seen in Fig. \ref{fig:dune_nd} (right panel). It will fulfill the role of TMS, measuring the momentum and sign of the charged particles exiting ND-LAr. Additionally, it will be able to measure neutrino interactions inside the HPgTPC, achieving lower energy thresholds than those of the ND and FD LArTPCs. It will also provide a uniform event acceptance, similar to the FD, which could not be achieved by ND-LAr + TMS. By doing so, ND-GAr will allow to constrain the relevant systematic uncertainties for the long-baseline analysis even further. A detailed discussion on the requirements, design, performance and physics of ND-GAr can be found in the DUNE ND CDR \cite{DUNE2021NDCDR}.

\section{Prototyping efforts}

The ProtoDUNE programme at CERN focuses on the development of large scale prototypes for the DUNE FD. After a successful operation of the ProtoDUNE-SP detector \cite{DUNE2021ProtoDUNESP}, a 700 ton single-phase LArTPC, a new iteration of detectors featuring the HD and VD designs has been commissioned. The recorded data will be used to validate the performance of the designs, perform calibration studies, and measure hadron-Ar interactions.

Among the prototypes for the DUNE ND, the 2x2 demonstrator at Fermilab stands out. It consists of four LArTPC modules with pixelated readout, installed in the MINOS ND cavern. The setup includes upstream and downstream tracking planes repurposed from MINERvA. Cooldown and filling of the detector finished in May 2024, and it has been operating in the NuMI beam since July. The goal is to demonstrate the reconstruction of beam neutrino events with a native 3D readout, in an environment with similar event rates to DUNE.

\section{Conclusions}

DUNE is a long-baseline neutrino oscillation experiment and neutrino observatory. It has potential to deliver ground-breaking results, like the unambiguous determination of the neutrino mass hierarchy and the discovery of leptonic CP violation. DUNE also has a rich programme on astrophysical neutrinos, and BSM both at the ND and FD. At the moment, there are active large-scale prototyping efforts at CERN and Fermilab, as well as a rich R\&D programme for DUNE Phase II detectors.

\bibliography{FML_DUNE_PIC2024.bib}{}
\bibliographystyle{JHEP}

\end{document}